\documentclass[publish]{nst}
\usepackage{amsmath}
\usepackage{subfigure,dcolumn}
\usepackage[T2A,T1]{fontenc}
\usepackage[russian,english]{babel}
\usepackage{epstopdf}
\usepackage{color}
\usepackage{hyperref}

\usepackage{listings}
\begin{document}

\title{Spin coating TPB film on acrylics and measurement of its wavelength shifting efficiency }\thanks{This work was supported in part by Guangdong Basic and Applied Basic Research Foundation under Grant No. 2019A1515012216. The work was also supported in part by NSFC grant (11505301) and Innovation Training Program for bachelor students at School of Physics in SYSU. Hang Yang and Zi-Feng Xu contributed equally to the project.}

\author{Hang Yang}
\affiliation{School of Physics, Sun Yat-sen University, Guangzhou 510275, China}
\author{Zi-Feng Xu}
\affiliation{School of Physics, Sun Yat-sen University, Guangzhou 510275, China}
\author{Jian Tang}
\affiliation{School of Physics, Sun Yat-sen University, Guangzhou 510275, China}
\author{Yi Zhang}
\affiliation{School of Chemistry, Sun Yat-sen University, Guangzhou 510275, China}

\email[Corresponding author, Jian Tang: ]{tangjian5@mail.sysu.edu.cn.}

\begin{abstract}
Scintillation light from liquid noble gas in a neutrino or dark matter experiment lies typically within the vacuum ultraviolet (VUV) region and might be strongly absorbed by surrounding materials such as light guides or photomultiplier. Tetraphenyl butadiene (TPB) is a fluorescent material and acts as a wavelength shifter (WLS) which can turn the UV light to the visible light around a peak wavelength of 425 nm, enabling the light signals to be detected easily for physics study. Compared with a traditional TPB coating method using vapor deposition, we propose an alternative technique with a spin coating procedure in order to facilitate the development of neutrino and dark matter detectors. This article introduces how to fabricate the TPB film on acrylics using the spin coating method, reports measurement of sample film thickness and roughness, shows the reemission spectrum, and quantifies the wavelength shifting efficiency (WLSE).
\end{abstract}


\maketitle

\section{Introduction}
Wavelength shifter (WLS) is critical in modern liquid noble gas detectors. It shifts ultraviolet light signal to visible light signal of particular wavelength. The wavelength of scintillation light form liquid noble gas is vacuum ultraviolet (VUV) varying from 80 nm to 200 nm. Light in this wavelength range would be strongly absorbed by most detector materials. Tetraphenyl butadiene (TPB) is among one of the favorite WLS options~\cite{Benson:2017vbw,Kuzniak:2018dcf,Poehlmann:2018sto} in a number of neutrino and dark matter experiments using liquid argon, such as MicroBooNE~\cite{Fleming:2012gvl}, DUNE~\cite{Abi:2018alz}, DEAP-3600~\cite{b15,b5}, DarkSide-20k~\cite{Aalseth:2017fik} and ArDM~\cite{Amsler:2010yp}, which absorbs UV light and re-emit lights in visible spectrum to be easily and effectively detected by photomultipliers tubes (PMT) or Silicon Photomultipliers (SiPM). WLS is also used in Cherenkov detectors to improve light yield~\cite{Dai:2008cp,Sweany:2011qh,Joosten:2016lcl}.
It is of important value for current and next-generation experiments to find a cost-effective way to coat the TPB on the surface of detector container.

A vapor deposition method~\cite{Bonesini:2018ubd} is commonly used in TPB film fabrication, and the reference~\cite{Howard:2017dqb} proposes a spraying method. Since the vapor deposition method asks for the high vacuum and spraying method has no control on the film thickness, we propose a fabrication of TPB film using the spin-coating method, which can act as an alternative option. The primary purpose of this study is to make TPB films using the spin coating method and measure the film geometry as well as its capability of shifting UV lights. The light shifting capability of the TPB material is quantum efficiency (QE), which is defined as average number of photons TPB reemits when it absorbs a single photon. QE is an intrinsic property of the material itself and is independent of the film condition, e.g. thickness and roughness. Since it is hard to obtain a single photon as incident light, it is difficult to measure QE directly.
Therefore, the shifting capability of the whole film under multiphoton incident light, which is called wavelength shifting efficiency (WLSE), is measured instead. WLSE turns out to be a result of folding QE with the film condition and the optical configuration. It is a more straightforward representation of TPB film performance in physics applications.

Although UV light from most of liquid noble gas has a wavelength below 200 nm and cannot even transport much longer distance in air or acrylic, the reference~\cite{Benson:2017vbw} shows a clear relationship of WLSE with incident light at different wavelengths. The WLSE at different wavelengths share similar trends when they vary with the film thickness~\cite{Francini:2013lua}. This enables us to perform our WLSE measurement at a selected wavelength of 254.5 nm, even without vacuum environment. 
TPB is known to form at least four polymorphic types of crystals, depending the deposition method~\cite{CEC2014,LPR2014}. The optical response at 250 nm including the absorption and in consequence the WLSE of these different types of TPB can be quite different~\cite{CG2010}. It is also known that the scinillation yield of macroscopic TPB crystals grown from solution is much higher than the yield of evaporated coatings~\cite{IEEE2009, Pollmann:2010gs}. Despite differences in the deposition method, it is assumed that the main results will still hold in the same way.

In this study, we first propose a different TPB coating technique in Sec.~\ref{sec:fabrication}. Then we describe the experimental setup to measure the optical properties from our TPB samples in Sec.~\ref{sec:method}. An analysis method follows, in Sec.~\ref{sec:model}, to quantify the WLSE with an emphasis on geometrical acceptance ratio corrections. In Sec.~\ref{sec:results}, we report the light reemission spectrum by our TPB samples and make a comparison of WLSE between our results and those reported in the recent study. Finally, we come to the summary and conclusion.

\section{Fabrication with Spin Coatings}
\label{sec:fabrication}

\begin{figure}[!htb]
	\centering
	\includegraphics[width=7cm]{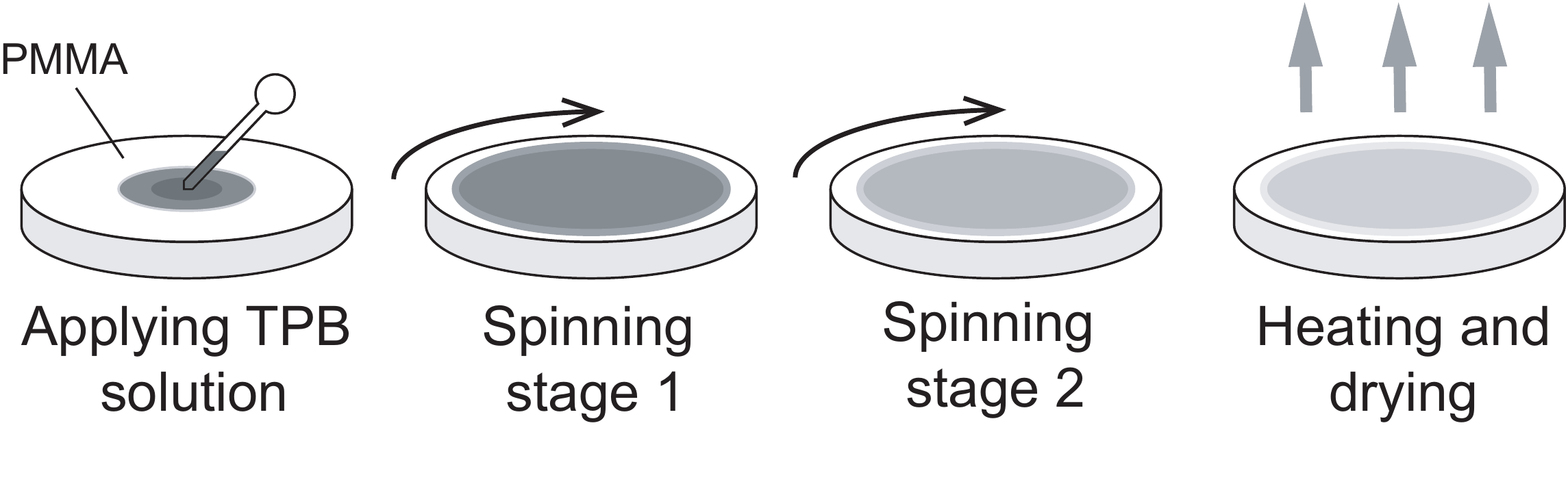}
	\caption{Spin coating procedure.}
	\label{spin}
\end{figure}

Without a loss of generality, we choose an acrylic disk as the base material for a demonstration of TPB coatings due to its high transparency in the interesting wavelength region. A description of the spin coating procedure is shown in Fig.~\ref{spin}.
Methylbenzene was used as the solvent for the TPB powder, {as its capability has been given in the literature~\cite{solvant}}.
According to the study in the literature~\cite{liyan}, the spinning speed was the major factor affecting the thickness of the film while the duration does not matter much.

We first sanded the surface of acrylics and applied the TPB solution with the methylbenzene as our solvent to the target surface. Then the TPB solution spreaded in stage 1 and the solvent methylbenzene volatilizes in stage 2, according to the previous study~\cite{b18,b21}. In this way, Stage 1 affects the thickness of the film the most as TPB was carried in the spilled solution. We performed preparative experiments to find the proper speed and the duration for stage 1. The solution barely spills in stage 2 due to its viscosity and the sample thickness would no longer change any more, which allows us to choose parameters for stage 2 based on experience without a loss of generality.

The solubility of TPB in methylbenzene was roughly around 0.021 g/ml by means of weight measurement in the room temperature. The concentration of 0.02 g/ml was chosen for the fabrication of samples.  The acrylic substrate in the round shape was fixed on the spinning-platform of the spin coating machine by a vacuum pump. TPB solution was added to the center of the acrylic disk manually. The spinning process consists of two stages with different speed and duration. Various combinations of speed and duration are used in a preparative experiment. The key parameters to prepare samples in the spin coating procedure are summarized in Table~\ref{pra}.

\begin{table}[!htb]
	\caption{Parameters.}
	\label{pra}
	\begin{tabular*}{8cm} {@{\extracolsep{\fill} } llr}
		\toprule
		Stage one & speed & 700 r/min \\
		 \hline
		  & duration & 6 s \\
		  \hline
		Stage two & speed    &  1000 r/min \\
		 \hline
		  & duration  & 20 s \\
		  \hline
		concentraion   &   & 0.02 g/ml \\

		\bottomrule
	\end{tabular*}
\end{table}
After the spinning procedure, the sample was placed on a heated table at 70$^{\circ}$C for 2 minutes to get dried. As an experience from experiment, the temperature and duration might not matter much as long as the film was completely dried.

\begin{figure}[!htb]
	\centering
    \includegraphics[width=8cm,height=6cm]{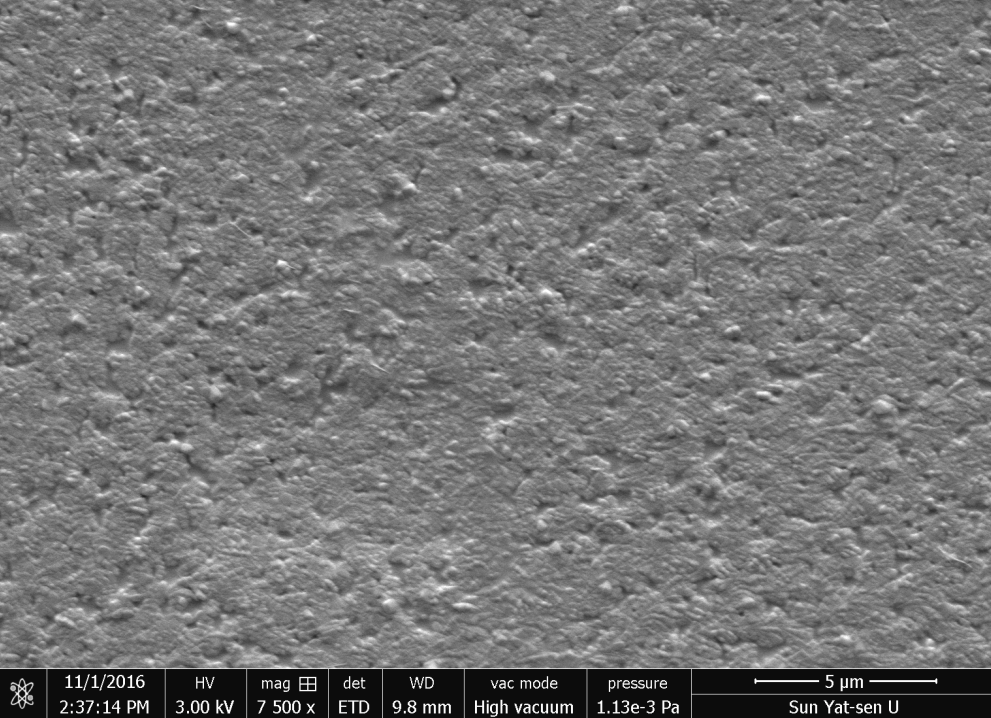}
	\caption{The surface profile measured by Scanning Electron Microscope (SEM) at the $5~\mu$m scale.}
	\label{zhaopian}
\end{figure}

The fabricated sample looks rather uniform and stable based on finger touches. Further surface measurements by SEM were performed and given in Fig.~\ref{zhaopian}~\footnote{The measurement was done in Nov. 1, 2019. The raw SEM picture had a wrong date due to the software.}. We saw tiny holes, which might be caused by the fast heating and drying processes. {The more uniformly the TPB film is fabricated, the better the fabrication method will be applied in the particle experiments. It deserves further investigation of temperature and cure to improve the surface roughness.} Samples were preserved in dark environment to prevent degradation~\cite{Chiu:2012ju,Jones:2012hm}.

We then measured the thickness and roughness of different TPB films made by the spin-coating method. A profilometer was used for our purpose. Each sample was scratched by a piece of metal to create a notch as the requirement of profilometer.

\begin{figure}[!htb]
	\centering
	\includegraphics[width=2cm,height=2cm]{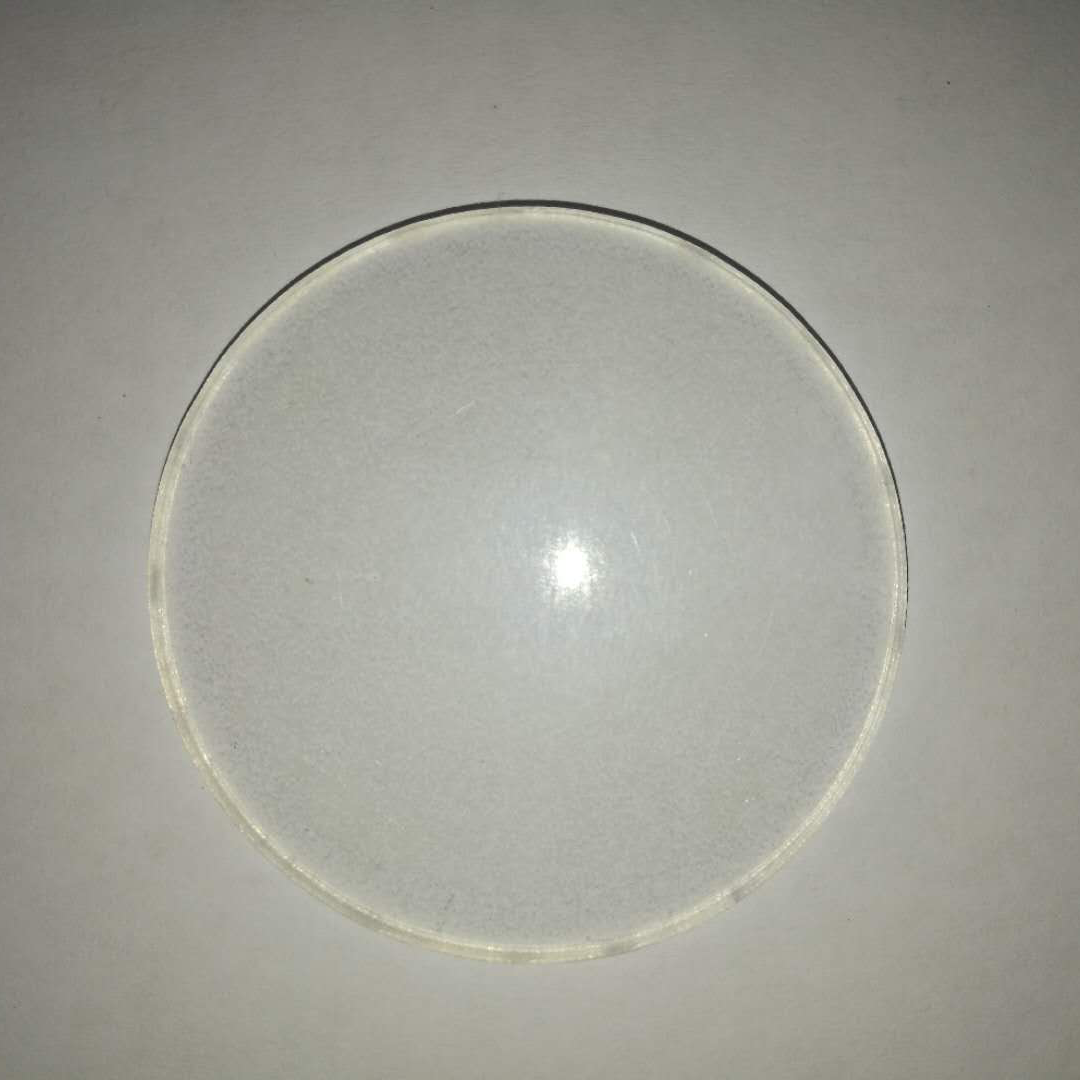}~~~~~~
	\includegraphics[width=3cm]{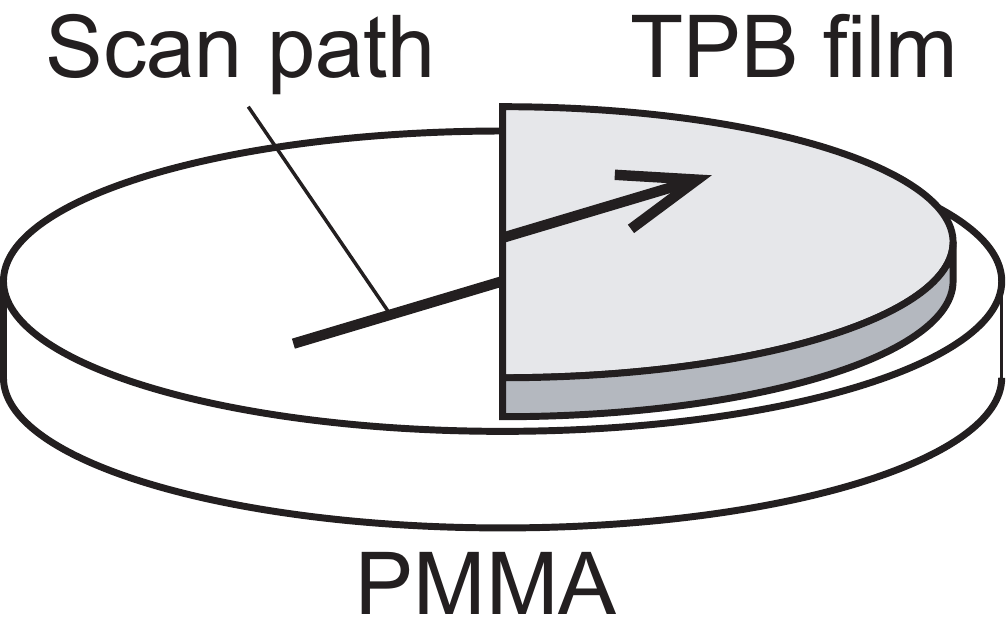}
	\caption{A photo of the fabricated sample on the left panel. The TPB half-coated samples and the scanning path by a profilometer on the right panel.}.
	\label{half}
\end{figure}
\begin{figure}[!htb]
	\centering
	\includegraphics[width=8cm]{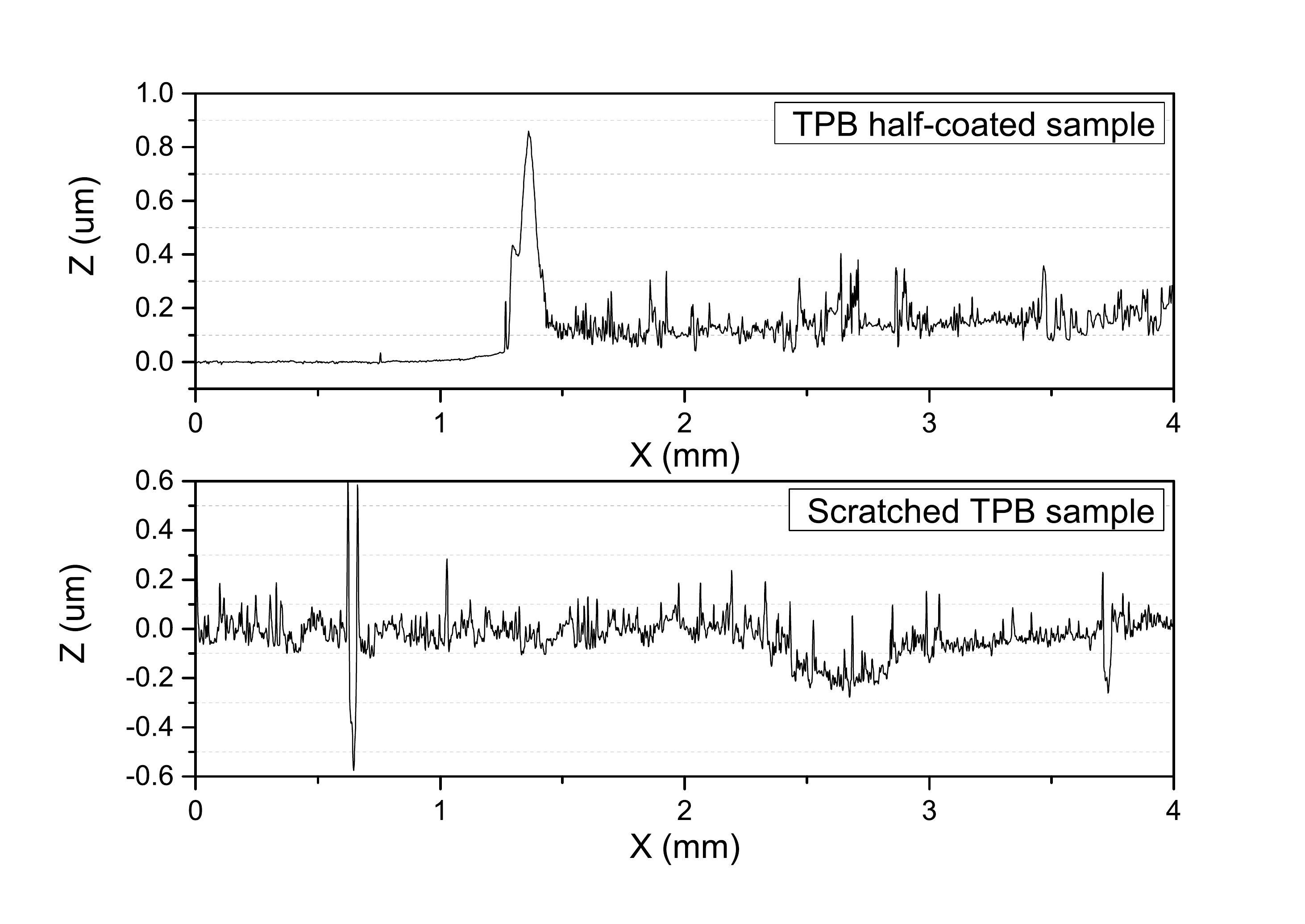}
	\caption{Results for the TPB half-coated and scratched samples by profilometer scans.}
	\label{taijie}
\end{figure}

The left panel in Fig.~\ref{half} is a photo of an acrylic disk coated with the TPB film. A special kind of samples, which have only half of their surface covered by the TPB film, shown in the right panel of Fig.~\ref{half}, were made for the purpose of roughness and thickness measurement. These special samples were made using the same parameter in the fabrication process except half of their surfaces were protected by a blank paper film before the TPB solution was added. The paper film was removed after the TPB solvent was completely dried. It was assumed that this half-pearl paper film would not affect the thickness and roughness of TPB film.
Fig.~\ref{taijie} shows scanning results of a TPB half-coated film and a scratched film done by a profilometer. Each scan covered a 4 mm long path and provided the height along the path.
It is then safe to consider the result on these paths representing the properties of the whole film because the film is almost uniformly distributed~\cite{b16}. The peak at the boundary in the TPB half-coated sample was caused by the protection film used in the fabrication process. We consider this accumulation as the result of wet edge between the solution and the final film, and assume that the edge does not affect the TPB-coated area far from the boundary. Thus this peak would be ignored in the following steps.

{In the result of a half-coated sample, the profilometer scan path covered an edge of the TPB film so that both surfaces of the TPB film and the acrylic substrate were included. The average height of the coated part with respect to the substrate was considered as the thickness and the uncertainties in a measurement of the unscratched part was considered as a representation of roughness. The measurements had been repeated several times, as results are shown in Table~\ref{rt_zonghe}. The scans for scratched samples used a path of 4 mm each time as well. The obvious valley represents the scratch we made on purpose. The difference between the bottom of the valley and the baseline is considered as the thickness of this sample.}
\begin{table}[!htb]
		\caption{Profilometer results. Roughness is represented by the standard error of thickness data.}
	\label{rt_zonghe}

	\begin{tabular*}{8cm} {@{\extracolsep{\fill} } lcr}
		\toprule
		measurements & thickness($\mu m$) &roughness($\mu m$)\\
		\midrule
		1  &    0.15    & 	0.051\\
		2  & 	0.17   & 	0.100\\
		3  & 	0.22   & 	0.074\\
		4  &    0.19   & 	0.081\\
		\hline
		average &0.18 & 0.076\\
		
		\bottomrule
	\end{tabular*}

\end{table}

As shown in Table.~\ref{rt_zonghe}, the thickness of TPB fabricated by the spin coating method was thinner than the reported result made by the vapor deposition. Therefore, it was expected that the WLSE here was slightly lower than the thicker sample produced by the vapor deposition. 

\section{Experimental Setup}
\label{sec:method}
\begin{figure*}[!htb]
	\centering
	\includegraphics[width=14cm]{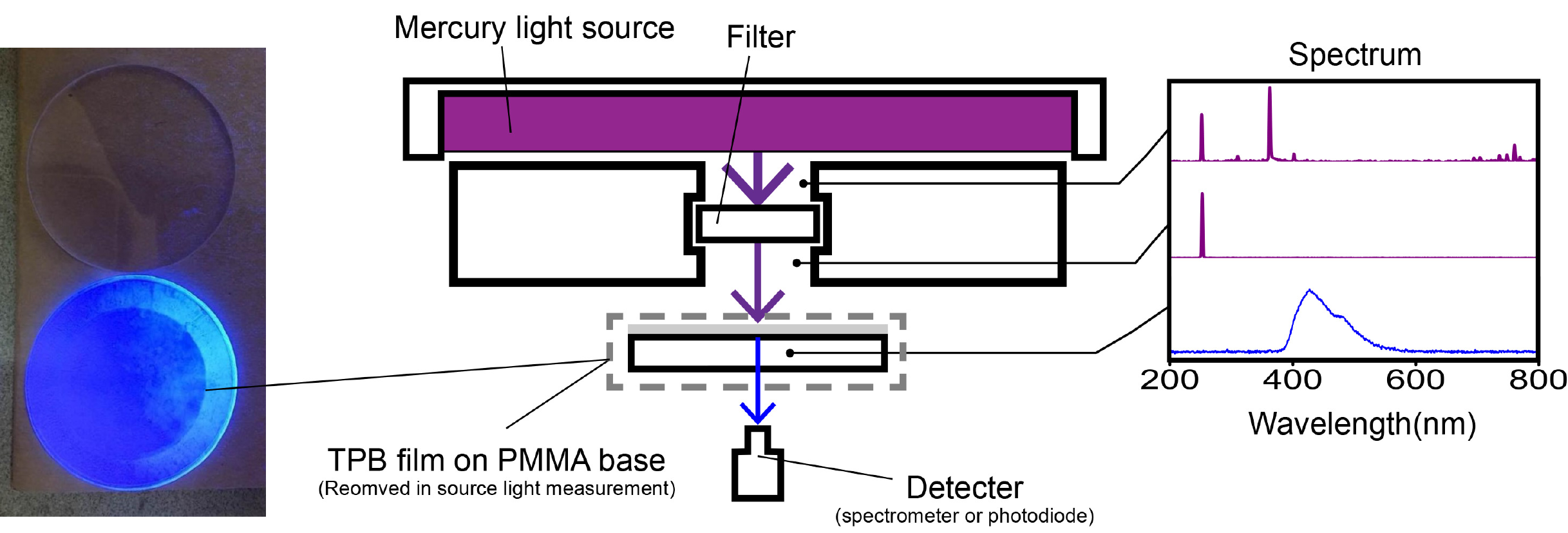}

	\caption{An experimental setup to measure reemission spectrum by spectrometer or photocurrents by photodiode.}
	\label{optic}
\end{figure*}
All apparatus used in experiment are listed in Table~\ref{yiqi}. A custom apparatus setup was built for measurements of spectrum and photocurrent. A schematics of this setup is shown in Fig.~\ref{optic}.
\begin{table}[!htb]
	\caption{A list of apparatus for optical measurements.}
	\label{yiqi}
	\begin{tabular*}{8cm} {@{\extracolsep{\fill} } lr}
		\toprule
		Apparatus & Type \\
		\midrule
		Spectrometer  & Ocean 2000\cite{ocean} \\
		Profilometer& KOSAKA ET150\cite{kosaka}    \\
		UV light   & WFH-204b\cite{wfh} \\
		The filter & Shengyakang\\\
		Si Photodiode   & LXD-66MQ  \\
		Spinner & KW-4A\cite{KW} \\
		Voltmeter & DM3000 Series\cite{dm}\\
		\bottomrule
	\end{tabular*}
\end{table}
We take a mercury lamp with filter installed right in front of the light exit window as the light source. The lamp provides lights ranging from 245 nm to 405 nm according to the characteristic spectrum of mercury. The spectrum of source light and reemission light overlaps at the wavelength of around 400 nm to 420 nm, which hinders our optical measurement. Therefore, a second filter, an interference filter with transmission peak at a wavelength of 254.5 nm was introduced to generate the monochromatic incident light.

From now on, the upper surface of filter or the TPB film are considered as light sources.
The spectrum measurements of both incident light and reemission light are done by a fiber spectrometer. The entrance of the fiber was placed at a fixed position and angle. Counting the photocurrent was done by a silicon photodiode. Due to the strong absorption of UV light in the air, the light sensor, i.e. the photodiode or the fiber entrance, was placed as close to the light source as possible. It means that the detector touches either the lower surface of filter or the acrylic substrate.

A DC power supply at 5 V was supplied to the Silicon photodiode set to work in the photoconductive mode, which was not compulsory but could slightly improve photocurrent measurements.
The photodiode had been calibrated by its vendor. This spectrometer covered the range from 200 nm to 800 nm. The fiber was supposed not to bring systematic uncertainties after a careful calibration. During data taking in each measurement, ambient noise in the lab environment was measured and subtracted with the help of a control and analysis software. In principle, two configurations in the optical measurement have slightly different geometries. The geometry-related correction have been considered in our analysis later.

\section{Analysis methods}
\label{sec:model}
WLSE is defined as the ratio between the reemission light intensity and the incident light intensity produced by the coated film. We first establish an optical model to help with the data analysis. For a monochromatic light at a wavelength of $\lambda$, each photon with energy of$\frac{\hbar c}{\lambda}$ would cause the photocurrent of $\frac{\hbar c R}{\lambda}$, where $R$ is the response of photodiode~\cite{handbook}. Hence, the number of photons $n$ can be calculated with the photocurrent in the following equation:
\begin{equation}
n\propto\frac{I}{\frac{h c R}{\lambda}}\,.
\end{equation}
Given a continuous light spectrum of $S(\lambda)$ (a normalized factor), this equation can be transformed into the following:
\begin{equation}
n\propto \frac{I}{\int d\lambda \frac{h c}{\lambda} R(\lambda) S(\lambda)}
\end{equation}

\begin{figure}[!htb]
	\centering
	\includegraphics[width=5cm]{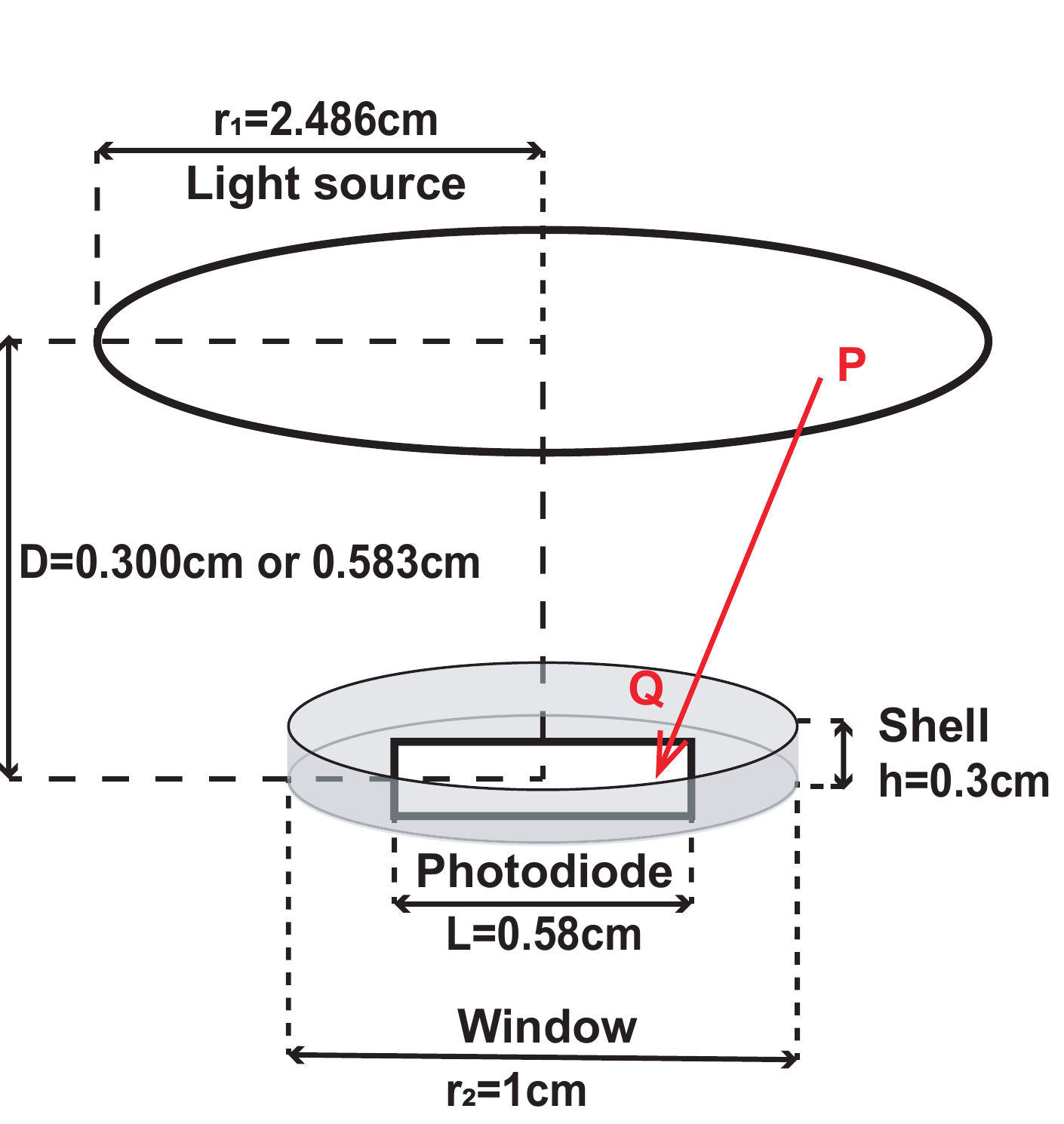}
	\caption{Simplified model for geometrical acceptance ratio calculation. D=0.583 cm for the reemission measurement and D=0.300 cm for the source light measurement.}
	\label{fig:moxing}
\end{figure}

Hence the WLSE would be:
\begin{equation}
\textrm{WLSE}=\frac{I_{\textrm{TPB}}}{I_\textrm{incident}}\times\frac{\int d\lambda'\frac{hc}{\lambda'} S_\textrm{incident}(\lambda') R(\lambda)}{\int d\lambda''\frac{hc}{\lambda''} S_{TPB}(\lambda'') R(\lambda)}\times \frac{1}{A}\,,
\label{eqn:WLSE}
\end{equation}
where A is a geometrical acceptance ratio which represents, roughly, the geometrical difference of apparatus configurations used to measure $I_{\textrm{incident}}$ and $I_{\textrm{TPB}}$. The numerical value for the ratio of two integrals in the middle of Eqn.~(\ref{eqn:WLSE}) is 0.96. In general, we need to measure photocurrent $I_\textrm{incident}$, $I_{TPB}$ using photodiode, and spectrum $S_\textrm{TPB}$, $S_\textrm{incident}$ by a fiber spectrometer.
In order to measure the photocurrent of the light source, on one hand, the TPB film was removed and the filter surface was considered as our light source. On the other hand, during a measurement of the reemission photocurrent, the TPB film was considered as the light source. The photodiode was placed as close to the light source as possible in each measurement. Thus the distance between a photodiode and a light source shall be determined by the thickness of the acrylics or filter. Since the filter was thinner, the Si photodiode received more light in light source measurement because it was indeed closer to the light source. A geometrical acceptance ratio was introduced to take the difference into account and make our measurements more accurate and self-consistent. This factor was determined by the geometry of apparatus configurations, which was simplified into the model shown in Fig.~\ref{fig:moxing} by means of geometrical optics.

{Among dimensions marked in Fig.~\ref{fig:moxing}, $r_2$ and $D$ were measured in the laboratory
with a precision of 0.01 mm while $L$ was provided by its vendor. These values will be taken into account during error propagations. Although the uncertainty in $L$ was not provided, we took a conservative uncertainty at 0.2 mm. The other parameters such as $r_1$ and $h$ were also measured by micrometer but ignored in the error propagation, because it turned out that their influence would be less than 0.1\% in our estimation. In a short summary, uncertainties of $r_2$, $D$ and $L$ were considered and would be processed for further quantitative analysis.}

Both source light and reemission light are considered as an even circular area Lambert source with the same size. The silicon photodiode has a square-shape sensitive area.
Suppose that P $(x_1,y_1)$ is a point on the area light source and Q $(x_2,y_2)$ is a point on photodiode. The intensity of light, measured by luminous flux, received by Q would decrease if Q is far from P or deviate from the exact front of P, for lambert light source, Q would be receiving luminous flux of:
\begin{equation}
dI_Q=B\frac{ \cos^2\theta }{r^2}dS_P dS_Q
\end{equation}
where $dS_P$, $dS_Q$ is a surface element near P and Q, respectively. B represents the brightness. Since we are interested in their ratio, it is safe to take the brightness B as 1.

{In terms of the reemission light, there are two major factors that could reduce the intensity of light. Firstly, there is a reflection when light enters the surface of the acrylic disk. Here we will ignore multiple reflections so that the reflected light will simply be lost. Secondly, light will be attenuated exponentially in acrylics. However, the acrylic substrate is very thin while the light attenuation length in acrylics at 420 nm is more than one meter according to the previous measurement~\cite{attenuation_length}. It means that the attenuation effect will be extremely weak. A simple estimate will tell us that the difference caused by an exponential attenuation should be less than 0.5\%. Therefore, the attenuation effect can be safely ignored here. In conclusion, an extra reflection factor has to be included to describe a reduction of light intensity due to reflections on acrylics. The transmission factor $T$ remains the same when swapping index 1 and 2 in Eqn.~(\ref{7}). Then $T$ will be simply factored in twice.}
\begin{equation}{\label{5}}
dI_Q=B\frac{ \cos^2\theta }{r^2}T^2dS_P dS_Q
\end{equation}

where
\begin{widetext}
\begin{equation}\label{7}
	T(\theta)=1-(\frac{1}{2}r_p^2+\frac{1}{2}r_s^2)=1-\frac{1}{2}[\frac{\tan(\theta_1-\theta_2)}{\tan(\theta_1+\theta_2)}]^2-\frac{1}{2}[\frac{\sin(\theta_1-\theta_2)}{\sin(\theta_1+\theta_2)}]^2	,\qquad		n_1\sin\theta_1=n_2\sin\theta_2
\end{equation}
\end{widetext}

\begin{figure}[!htb]
	\centering
	\includegraphics[width=8cm]{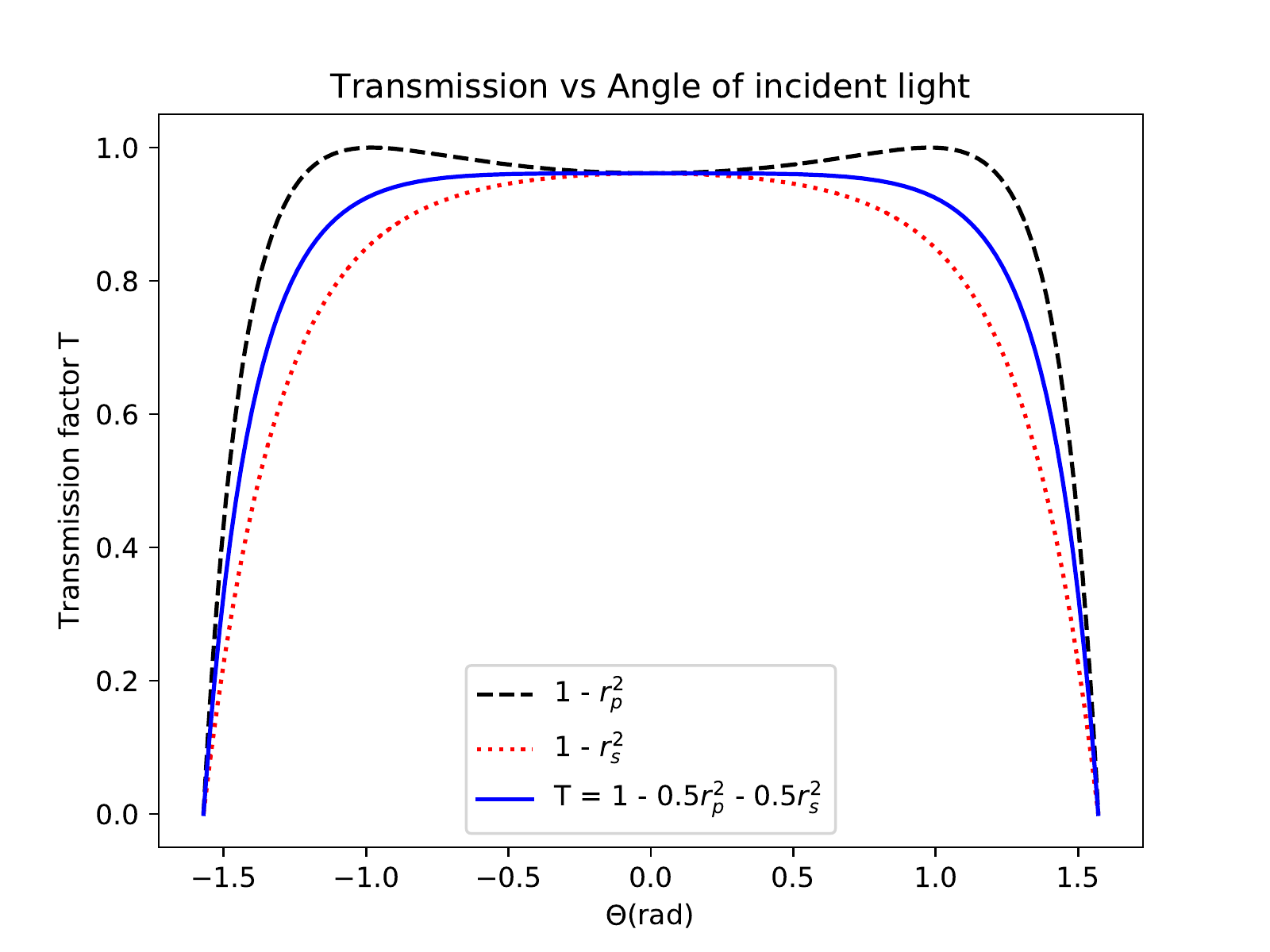}
	\caption{The transmission factor T varies with the angle of incident light.}
	\label{T}
\end{figure}

The transmission factor $T$ is calculated according to the Fresnel formula, assuming that incident light was not polarized there with equal contributions from P-wave(parallel) and S-wave(transverse). The calculation requires two refractive indices from the air and the acrylics: $n_1 = 1$ and $n_2 = 1.489$ based on the work~\cite{refractive_index}. Fig.~\ref{T} shows how the reflection factor varies with the angle of incident light.

It is necessary to integrate over the surface P and Q to get the ratio of light intensity at the source and the light intensity received by photodiode:
\begin{equation}\label{integral}
G=\frac{I_Q}{I_{all}}=\frac{1}{I_{all}}\iiiint\limits_{boundary} dI_Q
\end{equation}
Note that the transmission factor only appears in reemission light measurement. $I_{all}$ is the total intensity at the light source, which has the same value in two setups and thus would be canceled each other in the ratio. The boundary of ($x_1$, $y_1$) and ($x_2$, $y_2$) is circle and square respectively , with dimensions shown in Fig.~\ref{fig:moxing}. Parameter D is 0.300 cm for the source light measurement, and 0.583 cm for the reemission measurement. Note that the shell of photodiode could possibly block the light. This situation was cross checked and safely treated in the integral. This integral is then calculated numerically. The parameter G represents a light detection efficiency of photodiode. The geometrical acceptance ratio A for different setups is given by:
\begin{equation}
	A=\frac{G_\mathrm{reemission}}{G_\mathrm{source}}
\end{equation}
We immediately calculated results for $G_\mathrm{reemission} = 59.28$ and $G_\mathrm{source} = 69.86$. After that, we then obtain the geometrical acceptance ratio at 0.848. This value is smaller than 1, indicating that the photodiode is less likely to receive reemission light than source light, and this is mainly because the acrylic surface reflects the reemission light.
As the light source is much larger than the photodiode, this factor is not sensitive to a horizontal movement of the photodiode, which provides a high tolerance for the deviation of photodiode placement by hand. As mentioned above, the geometrical acceptance ratio A is determined by $r_2$, $D$ and $L$. Errors on measured values will be propagated in the results eventually.

\section{Results}
\label{sec:results}
\begin{table*}[!t]
	
	\caption{A list of uncertainties in measurements.}
	\label{uncertaintyTable}
	\begin{tabular*}{16cm} {@{\extracolsep{\fill} } llrrrr}
		
		\toprule
		Index(i)&Name  & Stdev($S_i$)& Precision($u_i$) & ($\partial f/\partial x_i$)& Measuring times\\
		\midrule
		1&	$U_i$ &8.394& 0.0001 mV&-0.00074 & 23 \\
		2&	$U_{i-dark}$ &0.075& 0.0001 mV&0.00074& 5\\
		3&	$U_r$ &1.881& 0.0001 mV&0.0033& 24\\
		4&	$U_{r-dark}$ &0.108& 0.0001 mV&-0.0033&6 \\
		5&	$r_2$  && 0.01 mm& 0.042& 1\\
		6&	$D$ && 0.2 mm& -0.049  & 1\\
		7&	$L$   &&0.2 mm& -0.0054&1\\

		\bottomrule
	\end{tabular*}
\centering
\end{table*}
\subsection{Photocurrent and spectra}
\begin{figure}[!htb]
	\centering
	\includegraphics[width=8cm]{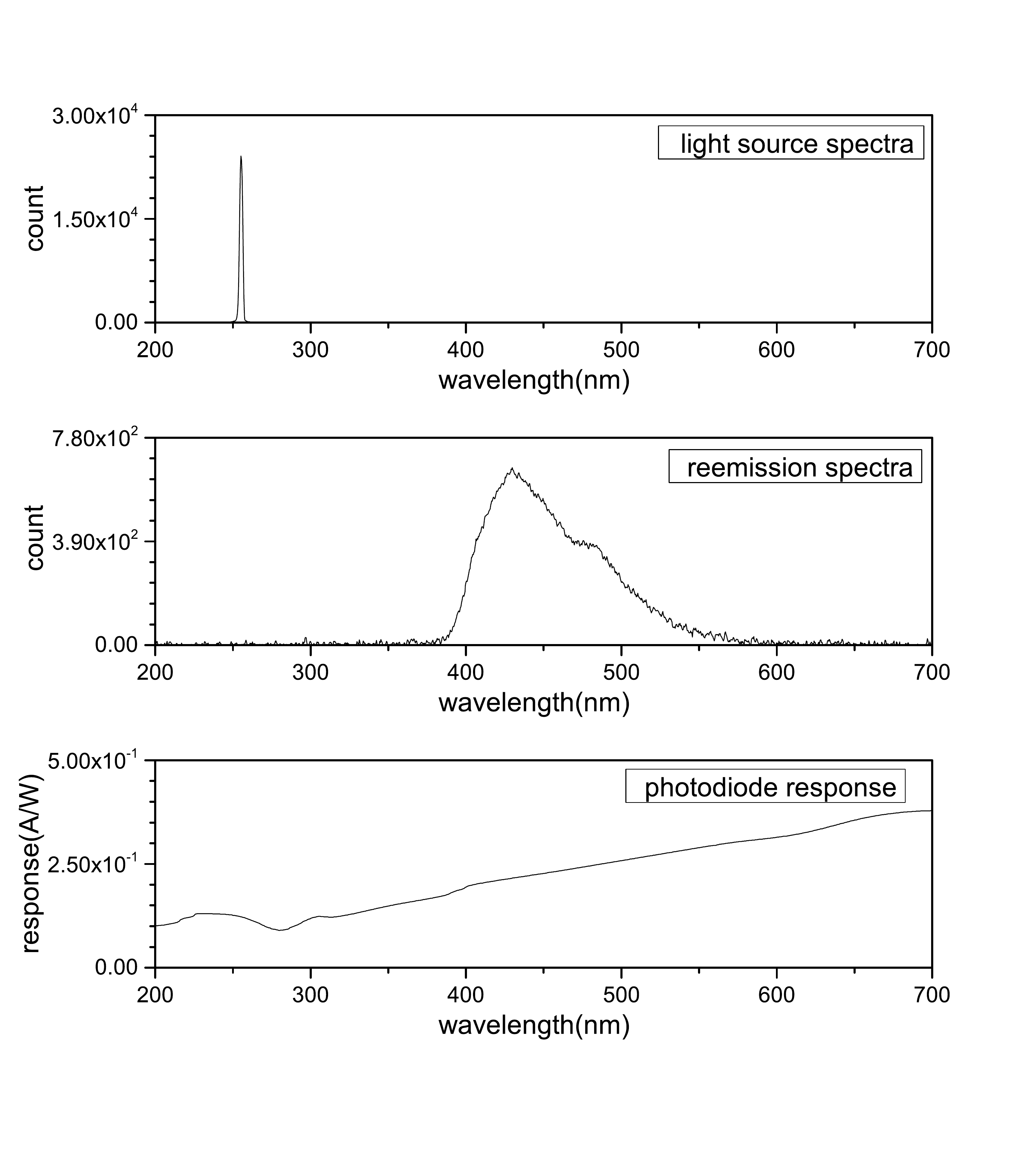}
	\caption{UV light source, TPB reemission spectrum, {the response spectra of the Silicon photodiode}.}
	\label{spe}
\end{figure}

The spectra of incident light and reemission light are shown in Fig.~\ref{spe}. The reemission spectrum matches the result reported in Ref.~\cite{Benson:2017vbw}, which indicates that the solvent and coating process would not affect TPB properties.

These spectrum will be used in calculation of WLSE.

A 100 k$\Omega$ resistor is connected to the photodiode in series to read out the photocurrent as voltage. {Note that this resistor value would be canceled in ratio so that it would not contribute to uncertainties in the end.} We actually replace the photocurrent with the generated voltage signal in Eqn.~(\ref{eqn:WLSE}) to avoid measuring the resistor which brings an extra uncertainty.

To guarantee validity of our results, photocurrent measurements of the light source and the TPB reemission were carried out in 6 times and 11 times, respectively. {UV light source was turned off between each measurement to avoid photodiode temperature rise which would further affect photocurrent.} Photocurrent is 341.8$\pm$8.4 mV labeled as $U_i$ in the source light measurement and 79.6$\pm$1.9 mV in the reemission measurement labeled as $U_r$ , where the error is standard deviation. Dark current is 4.1$\pm$0.1 mV labeled as $U_{i-dark}$ in the source light measurement and 4.8$\pm$0.1 mV labeled as $U_{r-dark}$ in the reemission measurement, respectively. Dark current is then subtracted in the analysis.

\subsection{Uncertainties}\label{uncertainty}

As mentioned above, we have to consider the uncertainty of WLSE propagated from voltage signals and geometric measurements.
For voltage measurements, on one hand, we have to include incident light voltage, reemission light voltage and their respective dark counts. The precision of voltmeter we used goes to the level of 0.0001 mV. {The statistical uncertainty comes from the standard deviation of light voltage with its own dark voltage subtracted. The systematic error is calculated according to the precision of our voltmeter. The partial derivatives WLSE with respect to these values will then be calculated analytically. For geometric measurements, on the other hand, we also use standard deviations as statistical errors and calculate systematic errors from the precision by micrometer. Note that these values take part in the numerical integral Eqn.~(\ref{integral}). Thus, the partial derivatives WLSE will be calculated numerically.  

We present a list of uncertainties in Table.~\ref{uncertaintyTable} which covers all values we measured with their precisions and standard deviations(Stdev). The notation for each value is defined there as well. If we define the combined statistical uncertainty $S_A$ with $v_A$ as its degrees of freedom and the systematic uncertainty $S_B$ with $v_B$ as its degrees of freedom, we can list the formula and results in error propagations as follows:
\begin{equation}
	S_A = \sqrt{\sum_{i=1}^{4}(\partial f/\partial x_i \times S_i)^2}
\end{equation}
\begin{equation}
v_A = \frac{S_A^4}{\sum_{i=1}^{4}\frac{(\partial f/\partial x_i \times S_i)^4}{v_i-1}}
\end{equation}
\begin{equation}
S_B = \sqrt{\sum_{i=1}^{7}(\partial f/\partial x_i \times u_i)^2}
\end{equation}
\begin{equation}
v_B = \frac{S_B^4}{\sum_{i=1}^{7}(\partial f/\partial x_i \times u_i)^4}
\end{equation}
\begin{equation}
S = \sqrt{S_A^2+S_B^2}
\end{equation}
\begin{equation}
v=\frac{S^4}{\frac{S_A^4}{v_A}+\frac{S_B^4}{v_B}}
\end{equation}
We check the coverage factor $t_p$ and the corresponding degrees of freedom $v$ from the t-distribution table~\cite{nist}. The uncertainty of WLSE can be obtained as $\Delta$WLSE = $t_pS = 0.044$ at the 95 \%~confidence level (C.L.).

\subsection{WLSE}
The result of WLSE based on our TPB samples is $0.251\pm0.044$ on average at 95 \%~C.L. This result is in line with the trend of measurements in Ref.~\cite{Benson:2017vbw}, where the WLSE of a 0.7~$\mu$m TPB film is around $0.4~\mu$m (thickness of our samples are around 0.18~$\mu$m shown in Table.~\ref{rt_zonghe}). Recalling that WLSE as a property of the film, is determined by intrinsic property of TPB material, QE, and the optical setup. It is, therefore, reasonable to expect that WLSE here would become lower as the film is thinner.

\section{conclusions}
We have successfully fabricated stable and well-functioning TPB films on acrylic disks by the spin coating method. We have measured those samples that thickness of the film is around 182 nm, and the surface was proven to be rather uniform by means of SEM. We have established an experimental setup to measure the optical properties of TPB films. We have checked TPB reemission spectrum, which perfectly matches the result reported in the previous work~\cite{Benson:2017vbw}. The WLSE in our samples has reached $25.1\pm4.4$\%  at 95\% C.L. as the similar level of TPB samples prepared by the vapor deposition method. The preliminary results show the feasibility of the spin coating techniques, though mass productions ask for more research and development. Tuning parameters in the spin coating procedure will likely increase the WLSE and meet requirements of different experiments. {One of shortcomings in the spin coating method remains how to deal with large panels without a round shape. Current commercial instruments cannot be applied any more. However, it is the relative velocity between the TPB liquid solution and the substrate to meet physics requirements. In the large scale applications, we might have to spill the liquid solution onto the surface by adapting well-designed jigs and heaters to fit the detector with a particular geometry. We will also have to elaborate the procedure in the clean room in the near future to avoid radioactive backgrounds which hinders the current technology in DM and neutrino experiments.} We expect this simple WLS coating technique to be optimized for future neutrino and dark matter detector constructions.

\section{Acknowledgement}
This work was strongly supported by the Center for Fundamental Physics Laboratory in SYSU. We appreciate great help from Prof. Han Shen and technicians in his team. Many thanks to Prof. Yue Zheng and Prof. Wen-Peng Zhu's help for sample measurements by SEM.


\end{document}